\documentclass[final,5p,times,twocolumn,authornumber]{elsarticle}

\usepackage[T1]{fontenc}
\usepackage[utf8]{inputenc}
\usepackage{amssymb}
\usepackage{amsmath}
\usepackage{microtype}
\usepackage[colorlinks=true,linkcolor=blue!50!black,citecolor=blue!50!black,urlcolor=blue!50!black]{hyperref}
\usepackage[nameinlink,noabbrev]{cleveref}
\biboptions{sort&compress}
\usepackage{amsthm}

\newcommand{\mpl}{M_{\rm Pl}}

\newcommand{\Lag}{\mathcal L}
\newcommand{\rd}{\mathrm d}

\newcommand{\bea}{\begin{align}}
\newcommand{\eea}{\end{align}}

\newcommand{\ddelta}{\Delta^{(J)}}

\journal{Physics Letters B}

\begin{document}
\begin{frontmatter}

\title{Gravitational baryogenesis beyond the spectator approximation}

\author[first,second]{David S. Pereira}\ead{djpereira@fc.ul.pt}
\author[first]{Beatriz A. Fernandes}\ead{biaaf@alunos.fc.ul.pt}
\author[first,second]{Jos\'e Pedro Mimoso}\ead{jpmimoso@fc.ul.pt}
\affiliation[first]{organization={Departamento de F\'isica, Faculdade de Ci\^encias da Universidade de Lisboa},
            addressline={Edif\'icio C8},
            city={Campo Grande},
            postcode={1749-016 Lisboa},
            country={Portugal}}
\affiliation[second]{organization={Instituto de Astrof\'isica e Ci\^encias do Espa\c co},
            addressline={Edif\'icio C8},
            city={Campo Grande},
            postcode={1749-016 Lisboa},
            country={Portugal}}

\begin{abstract} 
The standard gravitational-baryogenesis operator $\lambda\,\nabla_\mu R\,J^\mu$, with $\lambda\equiv \epsilon/M_\ast^{2}$, is usually treated as a spectator interaction that generates an effective chemical potential in a prescribed background. When included in the gravitational action, however, it defines a genuine curvature--matter-coupling variational problem, relevant for the baryon, lepton, and $B\!-\!L$ currents, whether described microscopically by particle-physics operators or macroscopically by a fluid current $J^\mu=n_Xu^\mu$. Up to a boundary term the interaction is equivalent to $-\lambda R\nabla_\mu J^\mu$, making its $f(R,{\rm Matter})$ character
manifest, but the metric equations remain open unless the metric dependence of $J^\mu$ is specified. For an arbitrary local realization $J^\mu(\Psi,g)$ we derive the universal part of the field equations and isolate the realization-dependent tensor generated by $\delta_g J^\mu$. In the vector-density realization the explicit $J^\alpha\nabla_\alpha R$ term cancels, but an algebraic term $-\lambda g_{\mu\nu}R\nabla_\alpha J^\alpha$ survives, so the theory admits only a partial effective-Planck-mass interpretation, $M_{\rm eff}^2=M_{\rm Pl}^2-2\lambda\nabla_\mu J^\mu$, and a time-dependent effective gravitational coupling during baryogenesis. Specializing to flat Friedmann-Lema\^itre-Robertson-Walker (FLRW) with a homogeneous current $J^\mu=n_Xu^\mu$, we obtain the modified Friedmann and Raychaudhuri equations, the associated continuity relation, and dimensionless diagnostics that quantify when the spectator approximation is controlled. We also discuss the implications for gravitational-baryogenesis studies in modified theories of gravity, providing a consistent General Relativity (GR) baseline for implementations in both
standard cosmology and modified-gravity frameworks.
\end{abstract}


\end{frontmatter}

%%%%%%%%%% INTRODUCTION %%%%%%%%%%%%%%%%%%%%%%%%%
\section{Introduction}\label{introduction}

Among the proposed mechanisms for explaining the baryon asymmetry, gravitational baryogenesis (GB)~\cite{Davoudiasl:2004gf} has drawn notable attention in recent years~\cite{Pereira:2024ddu,Pereira:2024kmj,Pereira:2025flo,Cruz:2025fuk,Whittingham:2026cbo,Goodarzi:2025npz,Oikonomou:2016jjh,Sooraki:2025jvi,Majeed:2025tjp,Sultan:2025uhu,Sultan:2025dqr,Alruwaili:2025gbw,Bhattacharjee:2020wbh,Odintsov:2016hgc,Sahoo:2019pat,Ramos:2017cot,Baffou:2018hpe,Bento:2005xk,Luciano:2022ely,Azhar:2020coz,Bhattacharjee:2020jfk,Nozari:2018ift,Usman:2024cya,Mishra:2023khd,Azhar:2021wvx,Shiromizu:2004cb,Agrawal:2021usq,Mishra:2024rut,Samaddar:2024qno,Samaddar:2024bnu,Jawad:2024uqo,Maity:2018bvh,Bhattacharjee:2021jwm,Hamada:2007vu,Rani:2022abc,Aghamohammadi:2017emk,Goodarzi:2023ltp,Atazadeh:2018xjo,Sultan:2024qpm,Rani:2023wco,Yang:2024gnf,Arbuzova:2019xti,Narawade:2024cgl,Alruwaili:2025uaz,Lambiase:2006dq,Lambiase:2007zz,DeFelice:2004uv,Sadjadi:2007dx,Saaidi:2010ey,Fukushima:2016wyz}. This mechanism is based on the dimension--six operator
\begin{equation}\label{eq:GBTerm} \Lag_{\rm GB}=\lambda\,\nabla_\mu R\,J^\mu, \qquad \lambda\equiv \frac{\epsilon}{M_\ast^{2}}, 
\end{equation} 
where $M_\ast$ is the EFT cutoff, $\epsilon$ is a dimensionless coupling constant, and $J^\mu$ is a current carrying the relevant global charge (typically $B$, $L$, or $B\!-\!L$). Normally this operator is interpreted as an effective chemical-potential source in a fixed cosmological background~\cite{Davoudiasl:2004gf}. This spectator\footnote{In the gravitational-baryogenesis literature the ``spectator approximation'' typically means that the operator $\lambda\nabla_\mu R\,J^\mu$ is used only as an effective source of a chemical potential, $\mu_{\rm eff}\sim \lambda\dot R$ (in an FLRW background), in the kinetic equations for the charge density, while its contribution to the metric field equations and the associated Bianchi-consistency relation is neglected. Equivalently, one assumes that the background evolution $H(t)$ and $R(t)$ is fixed by the GR equations for the dominant matter/radiation components and is not appreciably modified by the GB interaction.}viewpoint captures the baryon-bias mechanism, but it suppresses a second aspect of the same operator: Once inserted into the full action and varied with respect to the metric, it also changes the gravitational field equations. Specific realizations of this point were already emphasized by Arbuzova and Dolgov, who showed that the standard coupling can generate higher-order gravitational dynamics and strong instabilities, and later discussed how additional curvature terms may alter that conclusion~\cite{Arbuzova:2017zby,Arbuzova:2017vdj,Arbuzova:2016cem,Arbuzova:2023rri}. These results motivate a more systematic assessment of the gravitational sector implied by the operator~\eqref{eq:GBTerm}. In particular, once the interaction is promoted from a spectator source to an action-level coupling, it becomes natural to ask what the resulting metric field equations and consistency conditions are, and to what extent they depend on the microscopic realization of the current. The aim of this work is to provide such an action-level completion: we derive the universal operator-induced structure of the field equations, identify explicitly the realization-dependent contributions associated with the metric variation of $J^\mu$, and exhibit the corresponding cosmological backreaction implied by the Bianchi identity whenever the current is not conserved.

The paper is organized as follows. Section~\ref{sec:generic} derives the metric field equations resulting from the standard GB operator for a completely generic local current and makes explicit the separation between the universal operator-induced contribution and the realization-dependent term associated with the metric variation of the current. In Sec.~\ref{sec:realizations} we discuss representative realizations of $J^\mu$ and show how distinct variational prescriptions lead to different field equations, and therefore to different gravitational responses. We also summarize in this section the currents of direct relevance for baryogenesis and clarify how the formalism consistently encompasses both microscopic Noether currents and macroscopic fluid currents. In Sec.~\ref{sec:frw} we specialize the general equations to a spatially flat FLRW background, derive the modified Friedmann and Raychaudhuri equations together with the associated balance law, and provide a quantitative criterion for when the spectator approximation is parametrically controlled and can be considered. Section~\ref{sec:modified} comments on the implications of these results for gravitational-baryogenesis implementations in modified-gravity frameworks. Finally, Sec.~\ref{sec:discussionconclusion} discusses and summarizes the main results and outlines directions for further work.

%%%%%%%%%%% Variational completion %%%%%%%%%%%%%%%%%
\section{Variational completion}\label{sec:generic}

We begin with the action
\begin{equation}\label{eq:action}
S_{\text{Gravity}}=\int \rd^4x\,\sqrt{-g}\left[\frac{\mpl^2}{2}R+\lambda\,J^\mu\nabla_\mu R\right]+S_m[\Psi,g],
\end{equation}
where the matter fields are collectively denoted by $\Psi$, $S_m$ is the matter action defined using the matter Lagrangian, $L_m$, as
\begin{equation}
    S_m = \int \rd^4x \sqrt{-g} \,L_m\,,
\end{equation}
and the current is left arbitrary:
$J^\mu=J^\mu(\Psi,g)$.
Using
\begin{equation}\label{eq:ibp}
\nabla_\mu(RJ^\mu)=J^\mu\nabla_\mu R+R\nabla_\mu J^\mu,
\end{equation}
we may rewrite the interaction as
\begin{equation}\label{eq:actionibp}
S_{\text{Gravity}}\doteq \int \rd^4x\,\sqrt{-g}\left[\left(\frac{\mpl^2}{2}-\lambda  \nabla_\mu J^\mu\right)R\right]+S_m\,,
\end{equation}
where $\doteq$ means an equality up to a boundary term.
This rewriting is exact at the action level, but its variational content is preserved only if the metric dependence of $J^\mu$ is carried out consistently through the variation. From this rewriting we can extract two important results.
{First}, after integration by parts the GB term makes the coefficient of $R$ matter dependent, allowing one to define an effective Planck-mass function
\begin{equation}\label{eq:effPlanck}
M_{\rm eff}^2 \equiv M_{\rm Pl}^2-2\lambda\nabla_\mu J^\mu .
\end{equation}

This should be understood purely as an action-level reparametrization: the full field equations generally contain additional contributions from the metric variation of $\nabla_\mu J^\mu$, so the dynamics are not, in general, obtained by simply replacing $M_{\rm Pl}^2\to M_{\rm eff}^2$ in Einstein's equations (as will be discussed below). 

Furthermore, $M_{\rm eff}$ measures how strongly the GB operator modifies the gravitational sector at a given stage of the evolution. For example, the algebraic effect associated with the coefficient of the Einstein tensor becomes unavoidably important when
\begin{equation}
\frac{|M_{\rm eff}^2-M_{\rm Pl}^2|}{M_{\rm Pl}^2}
=
\frac{2|\lambda \nabla_\mu J^\mu|}{M_{\rm Pl}^2}
\sim 1,
\end{equation}
that is,
\begin{equation}
|\nabla_\mu J^\mu|\sim \frac{M_{\rm Pl}^2 M_\ast^2}{2|\epsilon|}.
\end{equation}

This condition does not determine the scale at which the GB operator is generated in the underlying effective theory, which is set by $M_\ast$; rather, it indicates when the GB contribution becomes dynamically relevant. In particular, it provides a criterion for identifying the epoch at which the GB-induced terms become comparable in magnitude to the standard gravitational terms in the background equations, as will be seen later. The corresponding temperature or energy regime is therefore model dependent.

Secondly, the theory is naturally viewed as a curvature--matter-{coupling}, $f(R,\text{Matter})$-type model\footnote{By $f(R,\mathrm{Matter})$ we mean, in a broad sense, a curvature--matter-coupled theory in which the gravitational sector depends not only on the Ricci scalar $R$ but also explicitly on matter variables.}~\cite{Harko:2010mv,Harko:2011kv,Harko:2024sea,Roshan:2016mbt,Katirci:2013okf,Harko:2015pma}, since the gravitational sector depends on matter through $\mathrm{J}\equiv\nabla_\mu J^\mu$. Importantly, the precise ``matter'' dependence is dictated by the realization of $J^\mu$; in particular, the model is not generically of the well-known $f(R,L_m)$ class~\cite{Harko:2010mv}.

Varying Eq.\eqref{eq:actionibp} with respect to $g^{\mu\nu}$ gives
\begin{align}\label{eq:deltaSgb}
\delta S &\doteq
\frac12\int \mathrm d^4x\,\sqrt{-g}\, \Big[
M_{\rm Pl}^2 G_{\mu\nu}
+2\lambda\Big(
(\nabla_\mu\nabla_\nu-g_{\mu\nu}\Box)\mathrm J \nonumber\\
&-\mathrm J R_{\mu\nu}
-\frac12 g_{\mu\nu}J^\alpha\nabla_\alpha R
\Big)-T_{\mu\nu}
\Big]\delta g^{\mu\nu} + \lambda \int \rd^4x \sqrt{-g}\, (\nabla_\alpha R) \delta_g J^\alpha,
\end{align}
where
\begin{equation}\label{eq:Tuv}
T_{\mu\nu}\;\equiv\;-\frac{2}{\sqrt{-g}}\,
\frac{\delta}{\delta g^{\mu\nu}}\!\left(\sqrt{-g}\,\mathcal L_m\right).
\end{equation}

The detailed metric variation of the gravitational-baryogenesis term is presented in~\ref{app:variation}. The term $\int \mathrm{d}^4x \sqrt{-g} (\nabla_\alpha R) \delta_g J^\alpha )$ corresponds to contribution from GB that depends on the explicit form of the current. To make that dependence explicit, we separate
\begin{equation}
\delta J^\alpha=\delta_\Psi J^\alpha+\delta_g J^\alpha,
\end{equation}
where $\delta_g J^\alpha$ is the metric-induced variation.
We then define
\begin{equation}\label{eq:DeltaJdef}
\lambda (\nabla_\alpha R)\,\delta_g J^\alpha
\equiv
\frac12\,\ddelta_{\mu\nu}\,\delta g^{\mu\nu}
+\text{(boundary terms)}.
\end{equation}
where $\Delta_{\mu\nu}^J$ encapsulates all the contributions to the metric field equations that $J^\mu$ can bring.   The full metric equations can therefore be written as
\begin{equation}\label{eq:FEgeneric}
{
\mpl^2 G_{\mu\nu}
+2\lambda\Big[(\nabla_\mu\nabla_\nu-g_{\mu\nu}\Box) \mathrm{J}-\mathrm{J}R_{\mu\nu}-\frac12 g_{\mu\nu}J^\alpha\nabla_\alpha R\Big]
+\ddelta_{\mu\nu}
=T_{\mu\nu}.
}
\end{equation}
or equivalently 
\begin{equation}\label{eq:FEgeneric1}
{M_{\text{eff}}^2} G_{\mu\nu}+(g_{\mu\nu}\Box-\nabla_\mu\nabla_\nu){M_{\text{eff}}^2}-{\lambda}g_{\mu\nu} \left(J^\alpha\nabla_\alpha R + R\mathrm{J}\right)
+\ddelta_{\mu\nu}
=T_{\mu\nu}.
\end{equation}

These results hold for an arbitrary current. They are an exact separation between a universal part, fixed by the operator itself, and a realization-dependent tensor $\ddelta_{\mu\nu}$, fixed only once the variational status of $J^\mu$ is specified. For a generic local current realization one may write
\begin{equation}\label{eq:genericdeltaJ}
\delta_g J^\alpha
=\mathcal A^\alpha{}_{\mu\nu}\,\delta g^{\mu\nu}
+\mathcal B^{\alpha\rho}{}_{\mu\nu}\,\nabla_\rho\delta g^{\mu\nu}
+\mathcal C^{\alpha\rho\sigma}{}_{\mu\nu}\,\nabla_\rho\nabla_\sigma\delta g^{\mu\nu}+\cdots,
\end{equation}
which implies
\begin{align}\label{eq:Deltageneral}
\ddelta_{\mu\nu}
&=2\lambda\Big[
(\nabla_\alpha R)\mathcal A^\alpha{}_{\mu\nu}
-\nabla_\rho\Big((\nabla_\alpha R)\mathcal B^{\alpha\rho}{}_{\mu\nu}\Big) \nonumber\\
&+\nabla_\rho\nabla_\sigma\Big((\nabla_\alpha R)\mathcal C^{\alpha\rho\sigma}{}_{\mu\nu}\Big)-\cdots
\Big].
\end{align}

To access the impact of the matter contributions induced by $\mathrm{J}$ we can use the Bianchi identity that provides a way to see the induced matter-geometry exchange.
Starting from the generic equation \eqref{eq:FEgeneric}, define
\begin{equation}
\Theta_{\mu\nu}
\equiv
2\lambda\Big[\nabla_\mu\nabla_\nu \mathrm{J}-g_{\mu\nu}\Box \mathrm{J}-\mathrm{J}R_{\mu\nu}-\frac12 g_{\mu\nu}J^\alpha\nabla_\alpha R\Big]
+\ddelta_{\mu\nu}.
\end{equation}
Then the field equations read $\mpl^2 G_{\mu\nu}+\Theta_{\mu\nu}=T_{\mu\nu}$, and therefore
\begin{equation}\label{eq:genericconsistency}
\nabla^\mu T_{\mu\nu}=\nabla^\mu \Theta_{\mu\nu}\; .
\end{equation}

As expected of a theory belonging to the class $f(R,\text{Matter})$ the stress tensor defined from $S_m$, Eq.~\eqref{eq:Tuv}, alone is not, in general, covariantly conserved. The gravitational sector depends explicitly on matter/current variables outside $S_m$, so the Bianchi identity enforces an exchange between the matter sector and the GB contribution. Moreover, this relation already shows that the matter equations and the choice of current realization are not independent. The usual spectator treatment corresponds to neglecting precisely these exchanges.

%%%%%%%%% Current realizations %%%%%%%%%%%%%%
\section{Current realizations}
\label{sec:realizations}

In the GB literature the current may be treated microscopically as a Noether current built from
matter fields (scalars or fermions), or macroscopically as a hydrodynamic charge flux in an
effective fluid description, and these choices need not induce the same $\delta_g J^\mu$
(e.g.\ \cite{Arbuzova:2017zby,Arbuzova:2017vdj}).

In this section we illustrate three representative and internally consistent realizations.
They should be viewed as variational choices: the same physical current may be represented
by different fundamental variables, and the realization-dependent tensor
$\Delta^{(J)}_{\mu\nu}$ in \eqref{eq:FEgeneric} encodes precisely this ambiguity.

\paragraph{(i) Fundamental contravariant vector (vector-field viewpoint)}
One may treat $J^\mu$ as a fundamental vector field whose components are held fixed under metric variation, i.e.\ $\delta_g J^\mu=0$. This is common practice when a vector is introduced as an independent field in the gravitational action, as in vector--tensor theories (a well-known example is Einstein--\ae ther theory, where a unit timelike contravariant vector is fundamental; see \cite{Jacobson:2000xp}). In this realization the metric-induced current variation vanishes and therefore
\begin{equation}
\Delta^{(J)}_{\mu\nu}=0.
\end{equation}
leads to the explicit $J^\alpha\nabla_\alpha R$ term in \eqref{eq:FEgeneric} to remaining active. Additionally, there is no clean reduction to an Einstein equation obtained solely by replacing $M_{\rm Pl}^2\to M_{\rm eff}^2$.

\paragraph{(ii) Fundamental covector (microscopic Noether-current viewpoint)}
In many microscopic realizations the natural object is the one-form current $J_\mu$.
For instance, in the scalar realization studied by Arbuzova and Dolgov the current is written with
a lowered index, $J_\mu\propto i(\phi^*\partial_\mu\phi-\phi\partial_\mu\phi^*)$, and the
contravariant current entering the action is obtained by raising the index with the metric
\cite{Arbuzova:2016cem}.
If $J_\mu$ is taken as fundamental and $J^\mu=g^{\mu\nu}J_\nu$, then
\begin{equation}
\delta_g J^\alpha=J_\beta\,\delta g^{\alpha\beta},
\qquad
\Rightarrow\qquad
\Delta^{(J)}_{\mu\nu}=2\lambda\,J_{(\mu}\nabla_{\nu)}R.
\end{equation}

Again the theory does not reduce to a pure scalar--tensor form, but the additional structure is
explicit and controlled, and it matches the expectation from microscopic currents whose index
structure is fixed before coupling to gravity \cite{Arbuzova:2017zby,Arbuzova:2016cem,Arbuzova:2017vdj}. The realizations discussed here are representative examples of possible variational prescriptions. They do not exhaust all microscopic constructions. In particular, fermionic currents in curved spacetime are naturally formulated in a tetrad formalism, and their metric variation can generate additional contributions beyond the simple covector prescription~\cite{Arbuzova:2017vdj}.

\paragraph{(iii) Fundamental vector density (fluid/Lagrangian-coordinate viewpoint)}
A particularly natural realization in macroscopic fluid descriptions is to take the
\emph{densitized} flux $\mathcal J^\mu\equiv\sqrt{-g}\,J^\mu$ as the fundamental variable.
This choice is in the spirit of standard variational formulations of relativistic fluids, in
which the fluid degrees of freedom are organized in terms of Lagrangian coordinates, velocity
potentials, or densitized currents, and the physically relevant flux is naturally represented
by a vector density \cite{Schutz:1970my,Brown:1992kc,Dubovsky:2011sj}.
At the same time, it is important to note that the classic Schutz/Brown constructions, as well
as modern non-dissipative fluid EFTs, are typically formulated for conserved currents.
In the present baryogenesis setting, where $\mathrm{J}\equiv\nabla_\mu J^\mu\neq 0$ during the
charge-violating epoch, the prescription $\delta_g\mathcal J^\mu=0$ should therefore be understood
as a phenomenological variational ansatz for the metric response of a coarse-grained charge flux,
unless one supplements the description with an explicit sourced/interacting fluid action
\cite{Iosifidis:2024ksa}.
This is sufficient for the purpose of deriving the bulk gravitational response associated with a
macroscopic current, but it should not be interpreted as a complete microscopic or off-shell fluid
completion.

In this realization one holds $\mathcal J^\mu$ fixed in the metric variation, $\delta_g\mathcal
J^\mu=0$, implying
\begin{equation}
J^\mu=\frac{\mathcal J^\mu}{\sqrt{-g}},
\qquad
\delta_g J^\mu=\frac12\,J^\mu g_{\rho\sigma}\delta g^{\rho\sigma},
\qquad
\Delta^{(J)}_{\mu\nu}=\lambda g_{\mu\nu}J^\alpha\nabla_\alpha R.
\end{equation}
This cancels the explicit $-\tfrac12 g_{\mu\nu}J^\alpha\nabla_\alpha R$ term inside
\eqref{eq:FEgeneric} and yields
\begin{equation}\label{eq:FEdensity} 
M_{\rm Pl}^2 G_{\mu\nu}
+2\lambda\Big[\nabla_\mu\nabla_\nu \mathrm{J}-g_{\mu\nu}\Box \mathrm{J}-\mathrm{J}R_{\mu\nu}\Big]
=T_{\mu\nu}.
\end{equation}
or equivalently (see \ref{app:density})
\begin{equation}\label{eq:FEvector} 
M_{\text{eff}}^2 G_{\mu\nu}
+\big(g_{\mu\nu}\Box-\nabla_\mu\nabla_\nu\big)M_{\text{eff}}^2
-\lambda g_{\mu\nu}R\mathrm{J}
=T_{\mu\nu}.
\end{equation}

For the non-conservation of $T_{\mu\nu}$, using the identity
\begin{equation}
\nabla^\mu\Big(\nabla_\mu\nabla_\nu \mathrm{J}-g_{\mu\nu}\Box \mathrm{J}\Big)=R_{\nu\alpha}\nabla^\alpha \mathrm{J},
\end{equation}
we find
\begin{align}
\nabla^\mu T_{\mu\nu}
&=2\lambda\Big[R_{\nu\alpha}\nabla^\alpha \mathrm{J}-\nabla^\mu(\mathrm{J}R_{\mu\nu})\Big] \nonumber \\
&=2\lambda\Big[R_{\nu\alpha}\nabla^\alpha \mathrm{J}-(\nabla^\mu \mathrm{J})R_{\mu\nu}-\mathrm{J}\nabla^\mu R_{\mu\nu}\Big] \nonumber \\
&=-\lambda \mathrm{J}\,\nabla_\nu R,\label{eq:bianchidensity}
\end{align}
where we used $\nabla^\mu R_{\mu\nu}=\tfrac12\nabla_\nu R$.

\paragraph{Conserved-current limit}
The conserved-current limit requires some care. In the usual gravitational-baryogenesis setting, the relevant charge current is not conserved while the charge-violating interactions are active, so that $\nabla_\mu J^\mu\neq 0$ during the baryogenesis epoch. After freeze-out, however, these interactions become inefficient compared with the Hubble expansion, typically in the sense that $\Gamma\ll H$, and one expects $\mathrm{J}\to 0$ up to residual washout effects. This late-time conservation is generally an \emph{on-shell} statement: it holds only after the matter dynamics are taken into account. It should be distinguished from the stronger case in which $\nabla_\mu J^\mu=0$ is imposed \emph{off shell}, namely as an identity or variational constraint on the admissible current configurations.

This distinction is important for the variational interpretation of the GB operator. If $\mathrm{J}=0$ only on shell, then the integrated-by-parts form of the interaction, $-\lambda R\,\mathrm{J}$, vanishes only after the matter equations are used. In that case, this rewriting does not by itself guarantee the disappearance of all bulk metric corrections, because the metric variation may still contain explicit terms involving $J^\alpha\nabla_\alpha R$ and/or the realization-dependent tensor $\Delta^{(J)}_{\mu\nu}$. By contrast, if $\mathrm{J}\equiv 0$ already off shell, then the interaction is a pure boundary term at the level of the variational problem and no bulk modification of the metric field equations remains.

The reduction to GR is therefore automatic only in realizations for which the field equations depend exclusively on $\mathrm{J}$ and its derivatives. In particular, for the vector-density realization, Eq.~\eqref{eq:FEdensity} reduces exactly to the Einstein equation when
$\mathrm{J}=0$. By contrast, in the contravariant-vector realization one obtains
\begin{equation}
M_{\rm Pl}^2 G_{\mu\nu} -\lambda g_{\mu\nu}J^\alpha\nabla_\alpha R = T_{\mu\nu},
\end{equation}
while in the covector realization one finds
\begin{equation}
M_{\rm Pl}^2 G_{\mu\nu} -\lambda g_{\mu\nu} J^\alpha\nabla_\alpha R
+2\lambda J_{(\mu}\nabla_{\nu)}R = T_{\mu\nu}.
\end{equation}

Hence GR is not recovered in general. Accordingly, the statement that the GB backreaction switches off once the current becomes conserved is realization dependent: current conservation is sufficient for bulk decoupling only in special realizations, most notably the vector-density one relevant for the macroscopic fluid treatment.

\subsection{Physically relevant currents in gravitational baryogenesis}\label{sec:currents}

The generic decomposition above is not an abstract luxury.
In actual gravitational-baryogenesis model building, the current is usually one of the physically meaningful global-charge currents,
\begin{equation}
J_B^\mu,\qquad J_L^\mu,\qquad J_{B-L}^\mu=J_B^\mu-J_L^\mu,
\end{equation}
or an effective current carrying the same quantum numbers.
At the microscopic level, these are built from particle fields.
For fermions one has schematically
\begin{equation}
J_X^\mu=\sum_i q_i^{(X)}\,\bar\psi_i\gamma^\mu\psi_i,
\end{equation}
with $X=B,L,B-L$ and with the appropriate charge assignments $q_i^{(X)}$.
Scalar realizations admit analogous Noether currents.
The analyses of Arbuzova and Dolgov are important precisely because they show, using explicit scalar and fermionic realizations, that the standard GB operator can feed back into the gravitational sector in a highly nontrivial way~\cite{Arbuzova:2017zby,Arbuzova:2017vdj,Arbuzova:2016cem}.

For cosmology, however, the same charge is often described after coarse graining by a hydrodynamic current
\begin{equation}
J_X^\mu=n_Xu^\mu+\nu_X^\mu,
\qquad \nu_X^\mu u_\mu=0,
\end{equation}
which reduces to the familiar thermodynamic form $J_X^\mu=n_Xu^\mu$ in a homogeneous FLRW background.
This is the level at which many baryogenesis computations are performed.
The central point of the present paper is that the action-level variation should be organized in a way that can cover \emph{both} descriptions.
The universal contribution is fixed by the GB operator itself, while the realization-dependent tensor $\ddelta_{\mu\nu}$ captures the fact that microscopic fermion, scalar, or fluid descriptions do not induce the same metric response.

This observation is also what makes the result useful.
A calculation performed with a prescribed background and a fluid current $J^\mu=n_Xu^\mu$ may be perfectly adequate in a deep spectator regime, but it should not be advertised as the generic gravitational completion of the operator.
Conversely, a conclusion drawn from a specific microscopic current, such as a fermion current, is not automatically universal.
Equation~\eqref{eq:FEgeneric} is therefore best read as a common action-level umbrella for the physically relevant currents used in gravitational baryogenesis.

%%%%%%%%%%%%%%% Cosmological implications: backreaction %%%%%%%%%%%%%%%%%%%%%%%%%%%%%%%%%
\section{Cosmological implications: backreaction}\label{sec:frw}
To model the cosmological backreactions induced by the GB term we will consider a flat FLRW metric 
\begin{equation}
\rd s^2=-\rd t^2+a(t)^2 \delta_{ij}\rd x^i \rd x^j,
\qquad H\equiv \frac{\dot a}{a},
\end{equation}
with a perfect-fluid matter source, $T^\mu{}_{\nu}={\rm diag}(-\rho,p,p,p)$. For the current we will adopt the vector-density realization because the current is being treated macroscopically as a hydrodynamic charge flux, $J^\mu=n u^\mu$, rather than as a specific microscopic Noether current. In variational formulations of relativistic fluids it is natural to take the densitized flux $\mathcal J^\mu\equiv\sqrt{-g}\,J^\mu$ as the fundamental variable (equivalently, to hold $\mathcal J^\mu$ fixed under metric variation), since it represents the comoving charge flux and can be defined without using the metric. This choice implies $\delta_g\mathcal J^\mu=0$ and therefore $\delta_g J^\mu=\tfrac12 J^\mu g_{\rho\sigma}\delta g^{\rho\sigma}$, yielding the simplified field equations used in Sec.\ref{sec:frw}. Other realizations, such as treating $J_\mu$ as fundamental as in explicit microscopic models, lead to different $\Delta^{(J)}_{\mu\nu}$ and therefore to different cosmological consequences. We therefore consider a homogeneous comoving current
\begin{equation}
J^\mu=n(t)u^\mu,
\qquad u^\mu=(1,0,0,0).
\end{equation}
giving
\begin{equation}\label{eq:Bfrw}
\mathrm{J}=\dot n+3Hn,
\end{equation} 
and yielding the Friedmann equation,
\begin{equation}\label{eq:Friedmann}
{
3\mpl^2H^2-6\lambda H\dot{\mathrm{J}}+6\lambda \mathrm{J}(\dot H+H^2)=\rho,
}
\end{equation}
and the spatial equation,
\begin{equation}\label{eq:Raychaudhuri}
{
-\mpl^2(2\dot H+3H^2)
+2\lambda\Big(\ddot{\mathrm{J}}+2H\dot{\mathrm{J}}-\mathrm{J}(\dot H+3H^2)\Big)=p.
}
\end{equation}

These equations are derived step by step in~\ref{app:frw}.
Both contain a derivative correction, $H\dot{\mathrm{J}}$, and an algebraic curvature correction, $\mathrm{J}(\dot H+3H^2)$, which is equivalently the surviving $\mathrm{J}R$ structure.

The continuity equation follows from the $\nu=0$ component of \eqref{eq:bianchidensity}, yielding
\begin{equation}
\nabla^\mu T_{\mu0}=-\big[\dot\rho+3H(\rho+p)\big],
\end{equation}
that translates into the modified continuity equation
\begin{equation}\label{eq:continuity}
{
\dot\rho+3H(\rho+p)=\lambda \mathrm{J}\dot R,
\qquad R=6(2H^2+\dot H).
}
\end{equation}

As expected, due to structural form of Eq.~\eqref{eq:bianchidensity}, the matter sector exchanges energy--momentum with the gravitational sector through a source term proportional to $\lambda\,\mathrm{J}\dot R$. The sign of $\lambda\,\mathrm{J}\dot R$ determines the direction of the exchange: for $\lambda\,\mathrm{J}\dot R>0$ the effective energy flow is into the fluid sector, whereas for $\lambda\,\mathrm{J}\dot R<0$ energy is transferred out of it. The magnitude and time dependence of this term therefore depend both on the background expansion history (through $\dot R$) and on the microphysical origin of the current nonconservation (through $\mathrm{J}$). The standard GR continuity equation is recovered whenever either $\mathrm{J}=0$ (current conservation) or $\dot R=0$ (constant curvature), in which case the exchange term vanishes identically.

\subsection{Backreactions}

A practical way to assess when the spectator treatment is self-consistent is to compare, term by term,
the GB-induced contributions in the background equations with the corresponding GR pieces.
In the vector-density realization, the $00$ and $ii$ equations can be written in the form
\begin{align}
3M_{\rm Pl}^2H^2
&=\rho \;+\;6\lambda H\dot{\mathrm{J}} \;-\;6\lambda\,\mathrm{J}(\dot H+H^2),
\label{eq:Friedmann_split}\\[2pt]
-M_{\rm Pl}^2(2\dot H+3H^2)
&=p \;-\;2\lambda\Big(\ddot{\mathrm{J}}+2H\dot{\mathrm{J}}-\mathrm{J}(\dot H+3H^2)\Big),
\label{eq:Raychaudhuri_split}
\end{align}
allowing one to isolate the standard GR part from the induced by the GB operator. The relative size of the corrections can then be captured by dimensionless ratios.
From Eq.~\eqref{eq:Friedmann_split} we define
\begin{equation}
    \delta_{\mathrm{J}}^{(F)}\equiv
\frac{|6\lambda\,\mathrm{J}(\dot H+H^2)|}{3M_{\rm Pl}^2H^2}
=\frac{2|\lambda\mathrm{J}|}{M_{\rm Pl}^2}\left|\frac{\dot H+H^2}{H^2}\right|.
\end{equation}
\begin{equation}
\delta_{\dot{\mathrm{J}}}^{(F)}\equiv
\frac{|6\lambda H\dot{\mathrm{J}}|}{3M_{\rm Pl}^2H^2}
=\frac{2|\lambda\dot{\mathrm{J}}|}{M_{\rm Pl}^2H},
\label{eq:deltaF_def}
\end{equation}
and from Eq.~\eqref{eq:Raychaudhuri_split}
\begin{equation}
\delta_{\ddot{\mathrm{J}}}^{(R)}\equiv
\frac{|2\lambda\,\ddot{\mathrm{J}}|}{M_{\rm Pl}^2|2\dot H+3H^2|},
\end{equation}
\begin{equation}
\delta_{\dot{\mathrm{J}}}^{(R)}\equiv
\frac{|4\lambda\,H\dot{\mathrm{J}}|}{M_{\rm Pl}^2|2\dot H+3H^2|},
\end{equation}
\begin{equation}
\delta_{\mathrm{J}}^{(R)}\equiv
\frac{|2\lambda\,\mathrm{J}(\dot H+3H^2)|}{M_{\rm Pl}^2|2\dot H+3H^2|}.
\label{eq:deltaR_def}
\end{equation}

These quantities provide an unambiguous diagnostic of backreaction: they measure the magnitude of
each GB-induced term relative to the corresponding GR scale in the same equation. The spectator regime corresponds then to the simultaneous smallness of these ratios,
\begin{equation}
\delta_{\dot{\mathrm{J}}}^{(F)},\ \delta_{\mathrm{J}}^{(F)}\ll 1,
\qquad
\delta_{\ddot{\mathrm{J}}}^{(R)},\ \delta_{\dot{\mathrm{J}}}^{(R)},\ \delta_{\mathrm{J}}^{(R)}\ll 1,
\label{eq:spectator_condition_full}
\end{equation}
so that the GB contributions remain parametrically subleading in both the constraint and evolution
equations. In this regime, one may consistently use the GR background solution as an external input
in the baryogenesis computation (e.g.\ in $\mu_{\rm eff}\propto \lambda\dot R$ and in the kinetic
equations), with corrections controlled by the maximum backreaction parameter
\begin{equation}
\delta_{\rm br}\equiv
\max\!\left\{
\delta_{\dot{\mathrm{J}}}^{(F)},\delta_{\mathrm{J}}^{(F)},
\delta_{\ddot{\mathrm{J}}}^{(R)},\delta_{\dot{\mathrm{J}}}^{(R)},\delta_{\mathrm{J}}^{(R)}
\right\}.
\label{eq:delta_br}
\end{equation}

Conversely, if $\delta_{\rm br}\gtrsim\mathcal O(1)$, the GB operator cannot be treated as a
spectator: it alters the background evolution appreciably and must be included self-consistently
together with the charge-violating dynamics that determine $\mathrm{J}$.

\subsection{Concrete nonconservation model}
A simple example can be obtained by parameterizing the current nonconservation as
\begin{equation}\label{eq:Gammasource}
\nabla_\mu J^\mu=\Gamma n,
\end{equation}
with constant rate $\Gamma$.
Then the current obeys
\begin{equation}\label{eq:nequation}
\dot n+3Hn=\Gamma n,
\end{equation}
so that
\begin{equation}\label{eq:Bmodel}
\mathrm{J}=\Gamma n,
\qquad
\dot{\mathrm{J}}=(\Gamma-3H)\mathrm{J},
\qquad
\ddot{\mathrm{J}}=\big[(\Gamma-3H)^2-3\dot H\big]\mathrm{J}.
\end{equation}

Substituting into \eqref{eq:Friedmann} gives
\begin{equation}\label{eq:FriedmannGamma}
{
3\mpl^2H^2-6\lambda H(\Gamma-3H)\mathrm{J}+6\lambda \mathrm{J}(\dot H+H^2)=\rho,
}
\end{equation}
or, equivalently,
\begin{equation}\label{eq:FriedmannGamma2}
3H^2(\mpl^2-2\lambda \mathrm{J})-6\lambda H(\Gamma-3H)\mathrm{J}+\lambda \mathrm{J}R=\rho.
\end{equation}

This form makes the non-conservation structure especially transparent. The backreaction ceases to be spectator-like not only when the derivative correction becomes large, but also when the algebraic curvature term $\lambda \mathrm{J}R$ becomes comparable to the Einstein piece. If $\Gamma\ll H$, both corrections can be typically tiny if the current is strongly diluted. If $\Gamma\sim H$, the derivative and algebraic corrections can be of comparable order and the standard spectator treatment loses parametric control. If $\Gamma>3H$, then $\dot{\mathrm{J}}>0$ and the exchange law \eqref{eq:continuity} shows that the operator can feed energy into the fluid whenever $\dot R>0$. This toy model is enough to show that the action-level completion carries nontrivial cosmological dynamics beyond the usual chemical-potential estimate.

\section{Implications for gravitational baryogenesis in modified gravity}
\label{sec:modified}

Gravitational baryogenesis has also been explored within a broad range of modified-gravity frameworks. A substantial body of work has investigated the implementation and phenomenology of this mechanism in such settings, including curvature--matter-coupled, scalar--tensor, and teleparallel extensions of GR, among others~\cite{Pereira:2024ddu,Pereira:2024kmj,Pereira:2025flo,Cruz:2025fuk,Whittingham:2026cbo,Goodarzi:2025npz,Oikonomou:2016jjh,Sooraki:2025jvi,Majeed:2025tjp,Sultan:2025uhu,Sultan:2025dqr,Alruwaili:2025gbw,Bhattacharjee:2020wbh,Odintsov:2016hgc,Sahoo:2019pat,Ramos:2017cot,Baffou:2018hpe,Bento:2005xk,Luciano:2022ely,Azhar:2020coz,Bhattacharjee:2020jfk,Nozari:2018ift,Usman:2024cya,Mishra:2023khd,Azhar:2021wvx,Shiromizu:2004cb,Agrawal:2021usq,Mishra:2024rut,Samaddar:2024qno,Samaddar:2024bnu,Jawad:2024uqo,Maity:2018bvh,Bhattacharjee:2021jwm,Hamada:2007vu,Rani:2022abc,Aghamohammadi:2017emk,Goodarzi:2023ltp,Atazadeh:2018xjo,Sultan:2024qpm,Rani:2023wco,Yang:2024gnf,Arbuzova:2019xti,Narawade:2024cgl,Alruwaili:2025uaz,Lambiase:2006dq,Lambiase:2007zz}. In these studies the baryogenesis estimate is often performed in close analogy with the GR spectator treatment: one adopts the modified-gravity background evolution, computes the effective chemical potential (typically proportional to $\dot R$ or its modified-gravity analogue), and then evolves the charge density using a kinetic or Boltzmann equation. The analysis developed here sharpens the interpretation of such computations by isolating an additional and logically independent effect: even in GR, once the operator $\lambda\nabla_\mu R\,J^\mu$ is included in the variational principle, it generically induces an exchange between the matter sector and the gravitational sector and can spoil the spectator approximation unless its contributions to the background equations remain parametrically subleading.

To make the scope precise, we stress that the explicit derivations in Secs.~\ref{sec:generic}--\ref{sec:frw} are carried out for the Einstein-Hilbert action supplemented by the standard gravitational-baryogenesis operator. This restriction is deliberate: it provides a clean GR baseline in which the backreaction generated by the GB operator itself can be isolated before introducing further complications in the gravitational sector. If the underlying gravitational action is enlarged, for example by additional curvature invariants or extra geometric degrees of freedom, then the background field equations are already modified before the GB operator is added. Nevertheless, once the interaction $\lambda \nabla_\mu R\,J^\mu$ is included at the action level and varied consistently, it should still generate additional contributions to the metric equations and modify the corresponding Bianchi-consistency relation. Thus, the main qualitative conclusion of the present work --- namely, that the GB operator need not behave as a pure spectator source once treated variationally --- is expected to persist beyond Einstein gravity, although the detailed equations and the relative importance of the new terms become theory dependent.

This observation strengthens rather than weakens the motivation for revisiting gravitational baryogenesis in modified gravity. Modified-gravity theories typically introduce new dynamical degrees of freedom (e.g.\ additional scalar, vector, or torsional modes) and/or explicit curvature--matter couplings, which already alter the background evolution and may already imply a nontrivial exchange law for $T_{\mu\nu}$. In such settings the GB operator should be expected to generate \emph{additional} backreaction channels beyond those intrinsic to the chosen gravitational theory. Consequently, if backreaction effects can invalidate a spectator treatment already in GR, they are \emph{a fortiori} expected to be present in modified gravity and may be even more consequential, because the extra degrees of freedom can source, amplify, or mediate the response to $\nabla_\mu J^\mu$.

This also suggests a practical bookkeeping principle for future work. In an explicit modified-gravity GB model one should distinguish at least two sources of departure from the usual spectator treatment: first, the change in $H(t)$ and $R(t)$ produced by the chosen theory of gravity; second, the additional backreaction produced by the GB operator when it is varied consistently. The two effects need not be interchangeable. In frameworks that already contain a curvature-matter coupling, the GB operator typically adds a second exchange channel rather than merely reparameterizing the first. For this reason, the consistency conditions derived are useful not only in GR but also as a preliminary check inside modified-gravity implementations of GB.

Concretely, a minimally consistent implementation of gravitational baryogenesis in a modified-gravity framework should therefore proceed in two steps. One first determines the background equations and the intrinsic conservation (or nonconservation) law implied by the underlying theory of gravity in the absence of the GB operator. One then adds the operator at the level of the action (or, equivalently, adds its contribution to the field equations derived from a consistent variational prescription), which introduces the additional realization-dependent tensor $\Delta^{(J)}_{\mu\nu}$ and modifies the Bianchi-consistency relation accordingly. Only after this second step can one decide whether the standard baryogenesis computation may be performed in a spectator regime. Operationally, this amounts to verifying that the GB-induced terms are parametrically small compared with the dominant modified-gravity contributions in the background equations, in direct analogy with the GR diagnostics in Sec.\ref{sec:frw}. If this is not the case, then the evolution of the modified-gravity degrees of freedom, the background geometry, and the charge-violating dynamics determining $\nabla_\mu J^\mu$ must be treated as a coupled system.

A related case is quintessence baryogenesis \cite{DeFelice:2002ir,Ahmad:2019jbm}, in which the curvature derivative is replaced by a scalar-field coupling of the form $\partial_\mu\phi\,J^\mu/M$. Up to a boundary term, this interaction may be rewritten as $-(\phi/M)\nabla_\mu J^\mu$, so that the nonconservation scalar again enters explicitly in the action. Although this does not lead to the same geometric reinterpretation discussed in the present work --- in particular, there is no analogue of the partial effective Planck mass $M_{\rm eff}$ and no $f(R,\mathrm{Matter})$-type structure --- the same general lesson applies: once the baryogenesis coupling is included consistently at the action level, it need not remain a pure spectator source and can induce backreaction on the background dynamics. In this sense, the present analysis is complementary to quintessential and scalar-driven completions of baryogenesis, where similar action-level effects can arise even though the source of the CP-violating bias is a scalar degree of freedom rather than $\nabla_\mu R$~\cite{Pereira:2024ddu}.

Moreover, the same lesson applies beyond the specific operator $\lambda\nabla_\mu R\,J^\mu$. Modified-gravity constructions and EFT parameterizations of the early Universe often introduce additional higher-dimensional operators similar to the GB one (for instance, couplings between curvature invariants $J^\mu$, couplings between matter scalars, such as $T$) which are then treated as spectator sources on top of a prescribed background. The present analysis indicates that such an assumption is not automatic: whenever an EFT operator is promoted from a purely phenomenological source term to an interaction included in the action, it generically contributes to the metric field equations, modifies the associated Bianchi-consistency relation, and can feed back on the background evolution. Therefore, in modified-gravity implementations it is advisable to apply an explicit spectator-consistency check not only to the GB operator itself but also to any analogous curvature--matter EFT term introduced in the model, by comparing its induced contributions to the dominant terms in the background equations in the same spirit as Eq.\eqref{eq:spectator_condition_full}.

\section{Discussion and conclusion}
\label{sec:discussionconclusion}

The gravitational-baryogenesis operator is most commonly employed in a spectator approximation, where it is interpreted as an effective chemical-potential source acting on a prescribed cosmological background. In this work we have shown that, once the same operator is treated as part of the gravitational action, it defines a nontrivial curvature--matter-coupled variational problem whose consequences depend in an essential way on how the current is realized. This point is particularly relevant for the currents used in practice in gravitational baryogenesis ($B$, $L$, and $B\!-\!L$ currents), which appear both in microscopic descriptions as Noether
currents of fundamental fields and in macroscopic treatments as hydrodynamic charge fluxes.

At the action level the interaction may be rewritten, up to a boundary term, as $-\lambda R\,\nabla_\mu J^\mu$, which makes explicit that the gravitational sector acquires an $f(R,{\rm Matter})$-type dependence through $\mathrm{J}\equiv\nabla_\mu J^\mu$ and motivates the definition of an effective Planck-mass function~\eqref{eq:effPlanck}.

Accordingly, one may speak of a \emph{partial} effective gravitational coupling $G_{\rm eff}\sim (8\pi M_{\rm eff}^2)^{-1}$ associated with the coefficient of $G_{\mu\nu}$. However, this identification is only a reparametrization of the action: the full metric equations are not obtained by a naive replacement $M_{\rm Pl}^2\to M_{\rm eff}^2$, because $\nabla_\mu J^\mu$ is not a metric-independent scalar and the variation of the operator contains additional realization-dependent contributions. In this sense, the GB backreaction is qualitatively distinct from more familiar curvature--matter couplings in which the effective gravitational coupling depends on $L_m$ or on the trace $T\equiv T^\mu{}_\mu$: here it depends on a nonconservation scalar and therefore on the reaction dynamics responsible for baryogenesis.

The exact variation naturally separates into a universal contribution fixed by the operator itself and a realization-dependent tensor $\Delta^{(J)}_{\mu\nu}$ that encodes the metric response of the chosen current variable. This provides a clean, model-independent framework for comparing different current implementations within a single variational completion. In the vector-density realization, motivated by standard variational formulations of relativistic fluids, the explicit $J^\alpha\nabla_\alpha R$ term cancels but a surviving algebraic term proportional to $g_{\mu\nu}R\,\mathrm{J}$ remains.

Specializing to flat FRW and a homogeneous current $J^\mu=n u^\mu$, we derived the modified Friedmann and Raychaudhuri equations together with the associated continuity law, and we proposed explicit dimensionless diagnostics to quantify when the GB-induced terms remain parametrically subleading. These criteria provide a practical test of the spectator assumption: when they are satisfied, the standard baryogenesis computation on a fixed background is controlled; when they are violated, the background evolution and the charge-violating dynamics determining $\nabla_\mu J^\mu$ must be treated self-consistently.

The conserved-current limit also requires some care. In standard gravitational-baryogenesis scenarios, the relevant current is generally not conserved while the charge-violating reactions are active, so that $\mathrm{J}\neq 0$ during the baryogenesis epoch. After freeze-out, however, one expects $\mathrm{J}\to 0$ up to residual washout effects. At the action level this implies that the integrated-by-parts form $-\lambda R\,\mathrm{J}$ vanishes \emph{on shell}. This does not by itself guarantee that all bulk metric corrections disappear in an arbitrary current realization, because the variation of the original operator may still retain explicit dependence on $J^\alpha\nabla_\alpha R$ and/or on $\Delta^{(J)}_{\mu\nu}$. The recovery of the Einstein equations is therefore automatic only in realizations for which the field equations depend exclusively on $\mathrm{J}$ and its derivatives. This is precisely what happens in the vector-density realization, for which the deviation from GR is transient and localized to the epoch in which baryogenesis is operative. By contrast, in more general realizations, explicit bulk terms can survive even when $\mathrm{J}=0$. Accordingly, the statement that the GB backreaction switches off after freeze-out should be understood as realization dependent rather than universal.

From a phenomenological perspective, the most direct consequences are associated with the early-time Hubble rate. If the charge-violating epoch were to persist to sufficiently low temperatures, a transient departure from the GR expansion history could in principle be constrained by standard probes of the pre-BBN or BBN-era expansion rate \cite{Will:2014kxa,Uzan:2010pm,Copi:2003xd,Alvey:2019ctk}. For the high-scale implementations usually considered in gravitational baryogenesis, such effects are expected to be suppressed once $\mathrm{J}\to 0$ well before nucleosynthesis. Additionally, although the present analysis is restricted to the homogeneous background, any transient GB-induced modification of the early expansion history may in principle affect observables sensitive to pre-BBN cosmology. In particular, nonstandard early-time expansion histories are known to leave imprints on primordial gravitational-wave backgrounds through their transfer functions \cite{Pettorino:2014bka,Assadullahi:2009nf}. Since the action-level completion introduces a time-dependent coefficient $M_{\rm eff}^2=M_{\rm Pl}^2-2\lambda\mathrm{J}$ in front of the Einstein tensor, it is also plausible that the tensor sector may be transiently modified while $\mathrm{J}\neq 0$. We stress, however, that the present work does not derive the quadratic action for perturbations or the tensor-mode propagation equation, and therefore no firm claim is made here about the size or observability of any direct SGWB signature. Such questions require a dedicated perturbative analysis and are left for future work.

Finally, the analysis supplies a useful GR-side baseline for gravitational-baryogenesis studies in modified-gravity frameworks. Since backreaction effects arise already in GR once the standard GB operator is varied consistently, they should generically be present --- and potentially amplified --- in theories with additional degrees of freedom or intrinsic curvature--matter couplings. A consistent implementation in modified gravity should therefore disentangle (i) deviations in the background evolution intrinsic to the chosen gravitational theory from (ii) additional departures induced by the GB operator itself, and analogous spectator-consistency checks should be applied to other EFT operators similar to GB introduced in such theories.

Overall, the results presented here sharpen the conceptual status of gravitational baryogenesis beyond the spectator approximation and provide a transparent variational framework in which microscopic and macroscopic realizations of the relevant currents can be compared. A natural next step is to solve the coupled system numerically in representative early-Universe histories with realistic charge-violating reaction networks, and to match the realization-dependent tensor $\Delta^{(J)}_{\mu\nu}$ to explicit microscopic current constructions, thereby connecting the action-level completion directly to particle-physics models of baryon and lepton number violation.

\section*{Acknowledgements}
We thank the anonymous referee for the valuable suggestions provided, which have greatly contributed to enhancing the quality of this manuscript. DSP and JPM acknowledge funding from the Fundação para a Ciência e a Tecnologia (FCT) through the research grants UIDB/04434/2020, UIDP/04434/2020 and PTDC/FIS-AST/0054/2021.
\appendix

\section{Variation of the integrated-by-parts form}\label{app:variation}

Starting from
\begin{equation}
S_{\rm GB}=-\lambda\int \rd^4x\,\sqrt{-g}\,R\mathrm{J},
\end{equation}
we vary
\begin{equation}
\delta S_{\rm GB}=-\lambda\int \rd^4x\,\delta(\sqrt{-g}R\mathrm{J}).
\end{equation}
Using
\begin{equation}
\delta\sqrt{-g}=-\frac12\sqrt{-g}\,g_{\mu\nu}\delta g^{\mu\nu},
\
\delta R=R_{\mu\nu}\delta g^{\mu\nu}+\big(g_{\mu\nu}\Box-\nabla_\mu\nabla_\nu\big)\delta g^{\mu\nu},
\end{equation}
we obtain
\begin{equation*}
\delta S_{\rm GB}
=-\lambda\int \rd^4x\,\sqrt{-g}
\Big[-\frac12 g_{\mu\nu}R\mathrm{J}\,\delta g^{\mu\nu}+\mathrm{J}\,\delta R+R\,\delta \mathrm{J}\Big].
\end{equation*}
After integrating by parts, the $\mathrm{J}\,\delta R$ term becomes
\begin{equation}
-\lambda\int \rd^4x\,\sqrt{-g}\,
\big[\mathrm{J}R_{\mu\nu}+g_{\mu\nu}\Box \mathrm{J}-\nabla_\mu\nabla_\nu \mathrm{J}\big]\delta g^{\mu\nu}.
\end{equation}
For the variation of $\mathrm{J}$ we use
\begin{equation}
\delta \mathrm{J}=\delta(\nabla_\mu J^\mu)=\nabla_\mu(\delta J^\mu)+\delta\Gamma^\mu_{\mu\alpha}J^\alpha,
\end{equation}
together with
\begin{equation}
\delta\Gamma^\mu_{\mu\alpha}=-\frac12 g_{\rho\sigma}\nabla_\alpha\delta g^{\rho\sigma}.
\end{equation}
Hence
\begin{equation*}
\begin{aligned}
-\lambda\int \rd^4x\,\sqrt{-g}\,R\,\delta \mathrm{J}
&=-\lambda\int \rd^4x\,\sqrt{-g}\,R\nabla_\mu(\delta J^\mu)\nonumber\\
&+\frac{\lambda}{2}\int \rd^4x\,\sqrt{-g}\,RJ^\alpha g_{\rho\sigma}\nabla_\alpha\delta g^{\rho\sigma}\\
&=\lambda\int \rd^4x\,\sqrt{-g}\,(\nabla_\mu R)\delta J^\mu +\nabla_\mu(R\delta J^\mu)\nonumber \\
&-\frac{\lambda}{2}\int \rd^4x\,\sqrt{-g}\,g_{\mu\nu}\nabla_\alpha(RJ^\alpha)\delta g^{\mu\nu}.
\end{aligned}
\end{equation*}
with $\nabla_\mu(R\delta J^\mu)$ being a boundary term. 

Since
\begin{equation}
\nabla_\alpha(RJ^\alpha)=J^\alpha\nabla_\alpha R+R\mathrm{J},
\end{equation}
the $R\mathrm{J}$ term cancels the contribution coming from $\delta\sqrt{-g}$, and we are left with
\begin{equation*}
\begin{aligned}
\delta S_{\rm GB}
&=\lambda\int \rd^4x\,\sqrt{-g}\,
\Big[(\nabla_\mu\nabla_\nu-g_{\mu\nu}\Box)\mathrm{J}-\mathrm{J}R_{\mu\nu}-\frac12 g_{\mu\nu}J^\alpha\nabla_\alpha R\Big]\delta g^{\mu\nu}\\
&+\lambda\int \rd^4x\,\sqrt{-g}\,(\nabla_\alpha R)\delta J^\alpha,
\end{aligned}
\end{equation*}
which is the result quoted in \eqref{eq:deltaSgb}.

\section{Vector-density realization and partial Jordan-frame rewrite}\label{app:density}

For the vector-density realization,
\begin{equation}
\mathcal J^\mu\equiv \sqrt{-g}J^\mu,
\qquad \delta_g \mathcal J^\mu=0,
\end{equation}
we have
\begin{equation}
\delta_g J^\mu=\frac12 J^\mu g_{\rho\sigma}\delta g^{\rho\sigma},
\qquad
\ddelta_{\mu\nu}=\lambda g_{\mu\nu}J^\alpha\nabla_\alpha R.
\end{equation}
Substituting into \eqref{eq:FEgeneric} removes the explicit $J^\alpha\nabla_\alpha R$ piece and yields \eqref{eq:FEdensity}:
\begin{equation}
\mpl^2 G_{\mu\nu}+2\lambda\big[\nabla_\mu\nabla_\nu  \mathrm{J}-g_{\mu\nu}\Box \mathrm{J}- \mathrm{J}R_{\mu\nu}\big]=T_{\mu\nu}.
\end{equation}
To rewrite this in terms of $M_{\text{eff}}^2=\mpl^2-2\lambda  \mathrm{J}$, note that
\begin{equation}
M_{\text{eff}}^2G_{\mu\nu}
+\big(g_{\mu\nu}\Box-\nabla_\mu\nabla_\nu\big)M_{\text{eff}}^2
=\mpl^2 G_{\mu\nu}+2\lambda\big[\nabla_\mu\nabla_\nu  \mathrm{J}-g_{\mu\nu}\Box  \mathrm{J}- \mathrm{J}G_{\mu\nu}\big].
\end{equation}
Since $G_{\mu\nu}=R_{\mu\nu}-\tfrac12 g_{\mu\nu}R$, this equals
\begin{equation}
\mpl^2 G_{\mu\nu}+2\lambda\big[\nabla_\mu\nabla_\nu \mathrm{J}-g_{\mu\nu}\Box \mathrm{J}-\mathrm{J}R_{\mu\nu}\big]
+\lambda g_{\mu\nu}R\mathrm{J},
\end{equation}
therefore
\begin{equation}
M_{\text{eff}}^2 G_{\mu\nu}
+\big(g_{\mu\nu}\Box-\nabla_\mu\nabla_\nu\big)M_{\text{eff}}^2
-\lambda g_{\mu\nu}R\mathrm{J}
=T_{\mu\nu},
\end{equation}
which is precisely Eq.\eqref{eq:FEvector}.

\section{FRW equations and continuity law}\label{app:frw}

In flat FRW,
\begin{equation}
G_{00}=3H^2,
\qquad G_{ij}=-(2\dot H+3H^2)a^2\delta_{ij},
\end{equation}
\begin{equation}
R_{00}=-3(\dot H+H^2),
\qquad R_{ij}=(\dot H+3H^2)a^2\delta_{ij},
\end{equation}
and for any homogeneous scalar $X(t)$,
\begin{equation}
\Box X=-\ddot X-3H\dot X,
\qquad
\nabla_0\nabla_0 X=\ddot X,
\qquad
\nabla_i\nabla_j X=-Ha^2\delta_{ij}\dot X.
\end{equation}
Applying these identities to \eqref{eq:FEdensity}, the $00$ component gives
\begin{equation}
\mpl^2G_{00}+2\lambda\big[(\nabla_0\nabla_0-g_{00}\Box)\mathrm{J}-\mathrm{J}R_{00}\big]=\rho.
\end{equation}
Now
\begin{equation}
(\nabla_0\nabla_0-g_{00}\Box)\mathrm{J}=\ddot{\mathrm{J}}-(-1)(-\ddot{\mathrm{J}}-3H\dot{\mathrm{J}})=-3H\dot{\mathrm{J}},
\end{equation}
so
\begin{equation}
3\mpl^2H^2-6\lambda H\dot{\mathrm{J}}+6\lambda \mathrm{J}(\dot H+H^2)=\rho,
\end{equation}
which is \eqref{eq:Friedmann}.
For the spatial components,
\begin{equation}
(\nabla_i\nabla_j-g_{ij}\Box)\mathrm{J}=a^2\delta_{ij}(\ddot{\mathrm{J}}+2H\dot{\mathrm{J}}),
\end{equation}
and therefore
\begin{equation}
-\mpl^2(2\dot H+3H^2)
+2\lambda\big(\ddot{\mathrm{J}}+2H\dot{\mathrm{J}}-\mathrm{J}(\dot H+3H^2)\big)=p,
\end{equation}
which is \eqref{eq:Raychaudhuri}.

Finally, from \eqref{eq:bianchidensity},
\begin{equation}
\nabla^\mu T_{\mu\nu}=-\lambda \mathrm{J}\nabla_\nu R.
\end{equation}
For $\nu=0$ one has
\begin{equation}
\nabla^\mu T_{\mu0}=-\big[\dot\rho+3H(\rho+p)\big],
\end{equation}
hence
\begin{equation}
\dot\rho+3H(\rho+p)=\lambda\mathrm{J}\dot R,
\end{equation}
which is \eqref{eq:continuity}.

\bibliographystyle{elsarticle-num}
\bibliography{biblio}

\end{document}